# Barriers to Using Static Application Security Testing (SAST) Tools: A Literature Review

Zachary Wadhams
Montana State University
Bozeman, Montana, USA
zacharywadhams@montana.edu

Clemente Izurieta
Montana State University
Pacific Northwest National Laboratory
Idaho National Laboratory
Bozeman, Montana, USA
clemente.izurieta@montana.edu

Ann Marie Reinhold
Montana State University
Pacific Northwest National Laboratory
Bozeman, Montana, USA
reinhold@montana.edu

## Abstract

Developers face a challenging problem with no clear solution. Modern software breaches can wreak havoc on businesses and individuals alike. With code vulnerabilities being a leading cause, securing applications must be a priority for developers. Static Application Security Testing (SAST) has the potential to harden applications by assisting in the identification and resolution of security vulnerabilities. Despite this, many development teams have not adopted SAST tools into their environment. In this paper, we survey the recent literature to uncover why some developers are apprehensive towards SAST and identify what specific problems they encounter when using it. We found a variety of usability problems developers face when using SAST. Some are inherent of the tool and ultimately require some level of developer investment while others are tool shortcomings that SAST tool creators must address. Ultimately, we argue that in order to drive widespread adoption and consistent SAST usage, developers will need to embrace that some investment is required. Simultaneously, developers will be more likely to integrate SAST tools into their workflows if the creators of SAST tools simplify many aspects related to tool usage. Surmounting the primary obstacles preventing the adoption of SAST requires full consideration of both the technical and human factors.

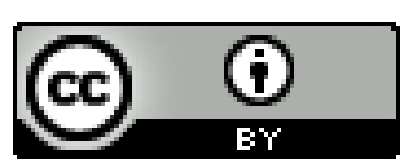

## 1 Introduction

Recent years have witnessed a surge in critical software security issues, impacting millions of people and causing billions of dollars in damages [1]. In July of 2024, a faulty CrowdStrike update unintentionally crippled Windows systems globally, highlighting the far-reaching consequences of software defects [6]. The 2020 SolarWinds attack stands as a well known example, where a code vulnerability allowed attackers to inject malicious code into software updates. This breach exposed sensitive data, disrupted critical infrastructure, and is estimated to have cost $100 billion to recover, impacting countless individuals and businesses [27]. These incidents underscore the critical need for robust secure coding practices throughout the software development lifecycle.

Fortunately, there are myriad of tools and techniques developers can take advantage of to secure their codebases. These include Dependency Scanning, Penetration Testing, Dynamic Code Analysis and Static Application Security Testing (SAST), among others[4, 14, 22]. Static Application Security Testing, in particular, is unique as it can be implemented at any time in the development lifecycle to identify and help to resolve vulnerabilities. It can achieve this because of how it works–by scanning source code without it needing to run or compile [31]. Such benefits have been recognized by Ayewah et al., who showed that many overlooked vulnerabilities were resolved after developers were alerted by warnings from SAST tools [3].

While implementing SAST is relatively easy and has shown benefits, some development teams still choose not to use it, opting for limited code analysis, focusing on security-critical components only, or relying mostly on manual code reviews [11, 16, 30]. This study aims to identify the root causes of developer apprehension regarding full SAST adoption. We focus on specific issues that discourage developers from initiating or abandoning SAST use. By understanding these challenges, we hope to shed light on areas for improvement in SAST usability and encourage further research directed at enhancing the developer experience.

## 2 Related Work

Previous surveys and studies have explored usability challenges associated with SAST tools. For instance, Johnson et al. [11] interviewed 20 software developers to understand their perspectives on usability issues. Charoenwet et al. [5] examined SAST tools for their effectiveness in security code reviews. In contrast, our study aims to gain a broad understanding of the current state of SAST usability through a comprehensive literature review.

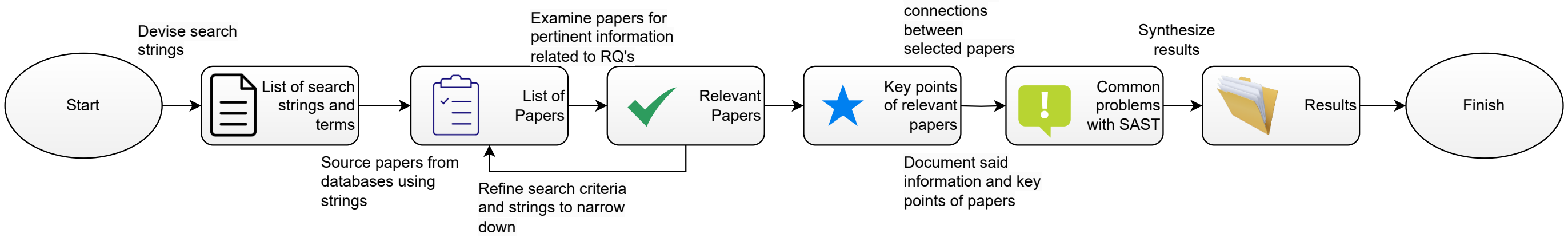

**Figure 1: Study Method Diagram. Search strings were devised and refined. Following refinement of strings, all 240 papers from the ACM Digital Library and all 108 papers from IEEE Xplore were examined for relevancy and either discarded or kept in accordance with our inclusion criteria. All relevant papers were read from cover-to-cover, thoroughly examined, and the issues developers encountered were cataloged. We then examined these issues systematically.**

## 3 Methodology

An overview of our study methodology can be found in Figure 1 and we will reference it throughout this section. The initial phase of our study involved choosing the databases from which to gather papers. We chose the ACM Digital Library and IEEE Xplore due to their extensive collections of peer-reviewed research articles and conference proceedings in the field of software security and development.

Next, we devised initial search strings as shown in the first step in Figure 1. These strings provided an estimate of the number of papers relating to our topic. Figure 2 displays these strings and the results of searching on them. The initial strings returned thousands of papers from each database. Consequently, we refined the strings, as detailed in Figure 2, until we reached a manageable number of papers. This is represented by the looping section from the clipboard to the check mark in Figure 1.

Our final search strings returned 240 papers from the ACM Digital Library and 108 papers from IEEE Xplore. Upon further examination of the results from the ACM Digital Library, we noted that the Brazilian Symposium of Systematic and Automated Software Testing [1] introduced 118 non-relevant papers into the 240 that were sourced. This inflation occurred because the acronym "SAST" was used in all publications from this conference, despite these papers not covering SAST. Removal of these conference papers reduced the number of ACM sources to 122 papers.

We then manually reviewed each of these 230 papers (108 from IEEE Xplore and 122 from ACM Digital Library sans the Brazilian Symposium of Systematic and Automated Software Testing papers) for their relevance. For the first pass, we specifically looked for papers that mentioned an implementation of SAST. We then realized that many papers mention static analysis or SAST in a general sense without any relation to the actual usage of these techniques. These papers were discarded. After the initial review, we identified a total of 89 relevant papers.

After identifying these papers, we analyzed them to identify key points related to SAST and developer usage, particularly the challenges encountered, shown at the star in Figure 1. We documented these problems and noted the number of papers in which they occurred.

[1] https://dl.acm.org/conference/sast

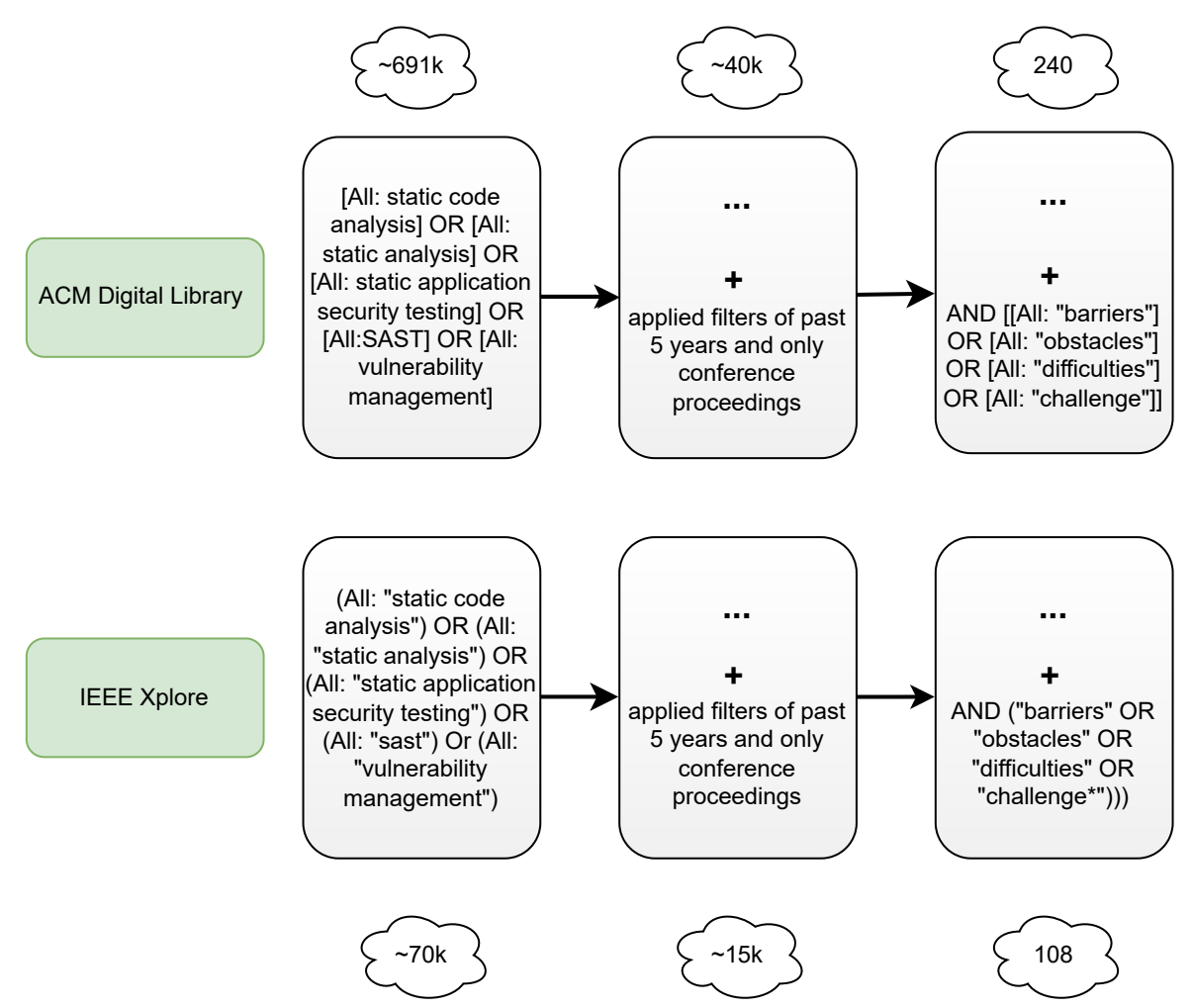

**Figure 2: Search String Refinement. This figure shows our search strings, refinement process, and the number of papers returned at each step of the process. We denote each refined string with "… +" to indicate that all the string everything after the + was concatenated to the string(s) indicated in the boxes to the left.**

## 4 Results and Discussion

In general, the number of publications increased annually from 2019 - 2023 (Figure 3). Except for a slight drop in 2022, there is a year-by-year increase in research on SAST implementation. This trend suggests that research interest in the use of SAST is increasing.

Many papers shared the following pattern. First, a tool would be selected and set up according to an organization's environment. These tools then require some level of attention to maintain which increases the effort needed from developers. As a result, SAST tools become cumbersome to use and thereby less attractive to those developers. Some papers suggest that these developers gradually reduced their use of the tools until they no longer served their original purpose and became defunct. Within organizations, the decline in usage can be attributed to several usability challenges (Figure 4), discussed in the following subsections.

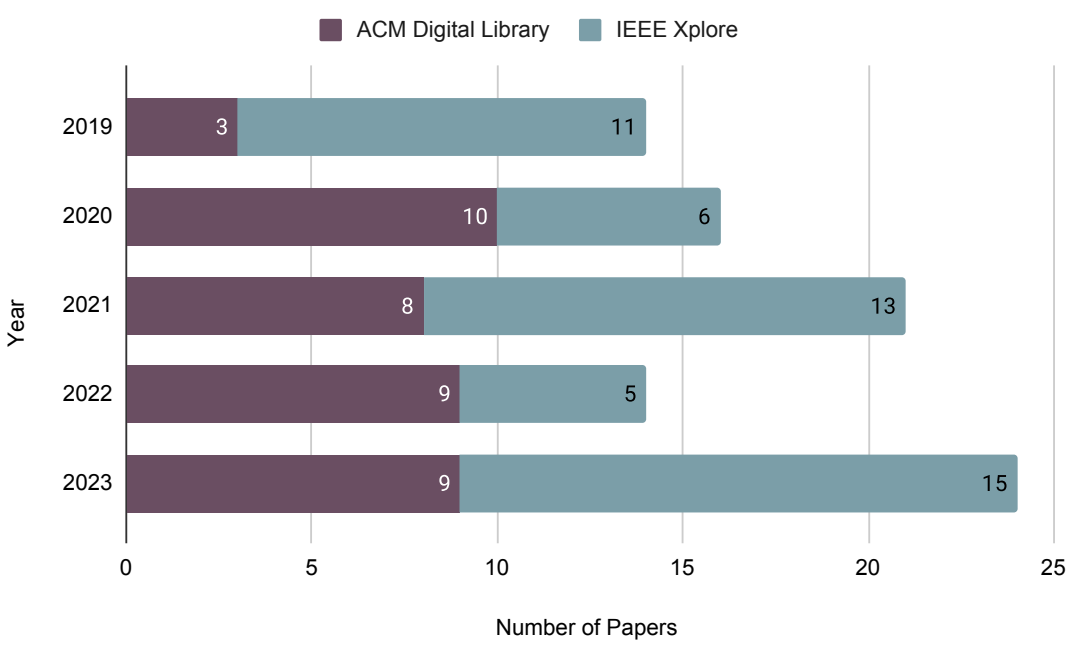

**Figure 3: Temporal Publication Trends in SAST Conference Proceedings. The ACM Digital Library is depicted in purple and IEEE Xplore in blue. An increasing trend over time can be seen, with the highest number of papers published in 2023.**

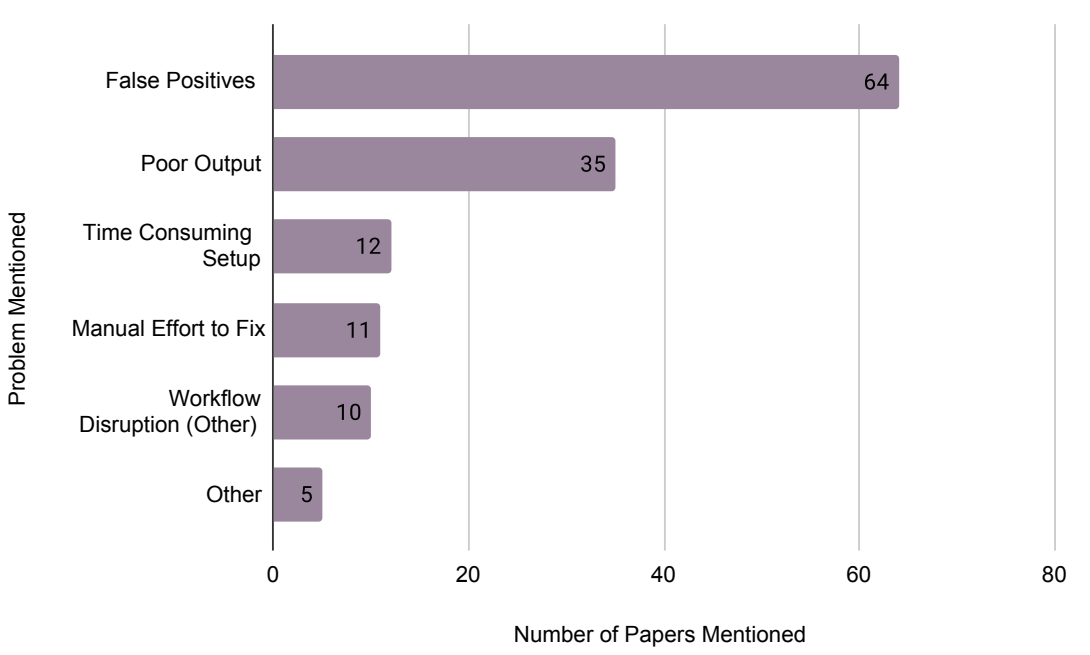

**Figure 4: Number of papers in which SAST related problems are mentioned. This figure shows the number of papers each issue was found in. False positives have nearly twice as many occurrences as the next highest complaint: poor output. Time consuming setup and manual effort to fix have a similar amount of instances. Workflow disruption (other) is a grouping of other workflow disruption issues. The final category, other, contains less frequently mentioned issues that were encountered.**

## 4.1 False Positives

In the context of SAST tools, a false positive refers to a situation where the tool incorrectly flags a piece of code as containing a security vulnerability or issue when, in fact, there is no real vulnerability present. Over two-thirds of papers cited false positives as a recurring pain point for developers (Figure 4).

These false positives can negatively impact developers in a variety of ways. A primary way this can manifest is by wasting the developer's time [23]. When a developer has to spend hours tracking down a reported issue only to find that it wasn't an issue, it can have a cascading result. This wasted time can lead to a decreased trust in the tool, general frustration, and overall reduced productivity [20].

The pervasiveness of false positives necessitates methods to mitigate them. Guo et al. states that it is imperative to implement strategies that minimize their occurrence and impact on the development process [9]. Such strategies are an open area of research.

The most common of these strategies is manual review, which requires a developer to investigate each reported issue to determine its validity [2]. If an issue is found to be a false positive, its identifier is added to an ignore list, and the SAST tool will no longer flag it.

## 4.2 Poor Output

Many SAST tools convey their results in a consistent format, either XML or JSON [11]. While this is useful for compatibility, developers are seldom interested in the sparse output and minimally formatted text that XML or JSON provides [20]. Depending on the size of the analysis files can contain thousands of lines and can be cumbersome and overwhelming [26].

A potential strategy to enhance the readability of outputs would involve constructing a parser that formats the issues. While a small number of tools have a built-in parser, the vast majority do not [28]. Adding a parser to facilitate consistent and familiar presentation has been shown to capture developer attention and potentially promote more consistent tool usage [28]. This additional effort could be regarded as addressing the next issue developers may encounter: time-consuming setup.

## 4.3 Time-Consuming Setup

While some SAST tools may be relatively simple to run-—e.g., just point them at a file structure—-others require an intensive setup that demands hours of investment [19]. This time investment may not be perceived as worthwhile, resulting in a lower adoption rate. A developer might spend hours attempting to configure a tool to run correctly within their environment, only to give up or deem it unworthy of the effort. Thus, time-consuming setup discourages uptake. We encourage the creators of these SAST to address this issue and–to the extent possible–provide clear, comprehensive instructions.

## 4.4 Manual Effort

When a developer uses a SAST tool, automatic code-fix suggestions may be appealing and encourage continued usage of the tool [21]. In contrast, when a tool alerts a developer without providing suggested fixes or a button for automatic correction, they can become frustrated by the manual effort required. This overall lack of guidance was a recurring theme relating to the manual effort required when using SAST tools.

A potential solution for the lack of guidance is offered by Linters, a static analysis tool that typically integrates into a developer's IDEs and provides real-time feedback [17]. Additionally, and as a bonus, many Linters have the auto-fix functionality that developers desire [10]. Thus, Linters have the potential to reduce the manual effort required.

### 4.5 Workflow Disruption

Static Application Security Testing can add time to a developer's workflow [20], potentially disrupting it. Disrupting a developer's workflow can have significant drawbacks, including decreased productivity, increased frustration, and potential resistance to adopting security practices [20]. When developers are interrupted by cumbersome tools, it can hinder their workflow and impede their progress on projects [20]. Additionally, disruptions may lead to context switching, where developers must shift their focus away from coding to address security issues, resulting in loss of momentum and increased cognitive load [8]. This can ultimately impact the quality and timeliness of software delivery. Therefore, minimizing disruptions to a developer's workflow is essential for maintaining productivity.

In order to achieve this, it is important for tools to operate swiftly, ideally to provide timely feedback on a developer's work without prolonged delays. Maintaining rapid deployment speed is paramount in the implementation of SAST [24]. Any delay in these deployments, no matter how brief, may lead to skepticism regarding a tool's effectiveness.

### 4.6 Other Problems

Various other one-off problems were mentioned in the corpus of papers we evaluated. The first of these is the lack of customizability of many tools [7]. This lack of customizability can manifest as a barrier in a few ways, with one being difficulty integrating smoothly with a development environment. With each environment being different and custom-tailored to the organization, these tools should strive to easily integrate with as wide of an audience as possible. The other way a lack of customizability can show up is in the actual running of the tool. Some may support a limited number of languages, scale poorly, or not allow users to customize their rules.

Another mentioned issue is unvalidated metrics [29]. Ideally, the results of a tool should be transparent and trustworthy, yet in many SAST tools, there is no level of proof to support them. Through the phenomena of "new version new answer", even something as simple as an updated tool version can give wildly different and unexpected results [25]. Because changes based on SAST results can have a significant impact, ensuring their accuracy would build trust in the tool and likely increase usage.

### 4.7 What's Next? Is It All Worth It?

Implementing SAST into a development environment helps secure code bases, and fosters more security-conscious developers [3]. Moreover, SAST tools can be efficient in finding and helping resolve security vulnerabilities [12, 15]. SAST tools can be viewed as investments, with the payout being a more robust and secure code base in a shorter time period than a manual review could provide. For instance, consider false positives. In the previously mentioned approach, detected false positives are reported to prevent them from appearing in future scans. With repeated implementation over time, a balance may be attained, with improved handling of false positives (e.g., databases containing known false positives). User investment in this effort ameliorates this problem.

Similarly, writing a parser to convert poor tool output to a better format can enhance a tool's usability [28]. While there was certainly a significant engineering effort involved, the firsthand developer feedback we received indicated a level of satisfaction with the tool that they did not have before the parser was implemented.

Some barriers to SAST implementation can be overcome through SAST creators developing auto-fix capabilities. Marcilio et al. [18] and Odermatt et al. [21] mentioned that auto-fix capabilities improved the overall usability and integration of tools into workflows and pipelines.

These observations suggest that both developers and SAST tool creators play independent but equally important roles in driving SAST adoption. This paper highlights the substantial advantages for developers who implement SAST proactively. In parallel, SAST tool creators must be responsive to user feedback to ensure continuous improvement in usability.

The successful implementation of SAST tools into development workflows incorporates technological advancements and is also heavily reliant on human factors. We know that the development community must invest significant time and effort to understand the tool's capabilities, interpret results, and address identified vulnerabilities. To further drive SAST adoption, a deeper understanding of the human element is needed. Further research should focus on identifying the institutional, social, and cognitive barriers that hinder SAST usage, such as skepticism about tool accuracy, or resistance to change. We assert that the future of the SAST field hinges on our ability to enumerate and address the human factors impacting developers.

## 5 Threats to Validity

This study offers valuable insights that contribute to the field of SAST and, by extension, secure software development. This section discusses potential threats to the validity of our study. By identifying these threats, our goal is to enhance the study's reliability and reduce the risk of bias.

Firstly, we followed a structured process for identifying relevant papers (Figure 1). Developing this process was the first action we took, therefore ensuring consistency through out each step of the literature review. Although we opted for a structured search strategy tailored to our research questions rather than a predefined protocol like Kitchenham's [13], this systematic approach ensured a rigorous selection of relevant literature. While our stringent search strings might have excluded some publications, we believe the resulting sample of 89 papers provides a sufficient window into the current state of the SAST field.

Finally, we opted to manually review all 89 selected papers instead of using automated approaches like Natural Language Processing (NLP). Some may claim that the lack of an automated aspect can limit the scalability of our approach. We argue, however, that this deliberate choice allowed us to gain a deeper understanding of the breadth of the problems by directly engaging with the authors' ideas. Our focus was on comprehending the range of issues, rather than the granular details explored in each paper. The manual review also facilitated our discussion of potential solutions. By directly encountering the developers' words, we were better equipped to formulate our thoughts on addressing the challenges they faced.

## 6 Conclusion and Future Work

This paper delivers a unique two-fold contribution to the Software Engineering and Cyber Human Factors communities. It dives deep into SAST usability challenges, offering valuable insights directly relevant to both SAST tool creators and development teams currently using or considering these tools.

Tool creators can take our results to better inform their development of new and improved SAST tools. Specifically, our findings on developer needs and challenges related to SAST usability can guide the design of more user-friendly interfaces and output display, improved false positive reporting mechanisms, simplified setup, and automatic fixing features.

Development teams can also leverage our findings to gain insights into potential challenges during SAST implementation. This knowledge can inform their SAST planning process when it comes to false positives, improving tool output and accounting for workflow disruption, ultimately empowering them to use these tools more effectively.

While this study focuses solely on the breadth of how these problems are mentioned, examining the depth at which specific papers discuss them could be valuable future work. This could involve investigating the depth of discussion for each problem and potentially inferring the severity or overall impact it poses for developers. A study of this kind could benefit from some form of NLP to assist in identifying key details in text that signal severity. Our review may also be expanded upon by extending the search back another 5 or 10 years. This could potentially help to gain a broader understanding of the uncovered issues and also identify how their prevalence has changed over time.

## Acknowledgments

This research is supported by TechLink (TechLink PIA FA8650-23-3-9553). The ChatGPT Large Language Model was used in this paper for spell-checking and grammatical enhancements. A full list of our 89 final papers this review was sourced from can be found by following this **link.**

## References

[1] Rahaf Alkhadra, Joud Abuzaid, Mariam AlShammari, and Nazeeruddin Mohammad. 2021. Solar Winds Hack: In-Depth Analysis and Countermeasures. *2021 12th International Conference on Computing Communication and Networking Technologies (ICCCNT)*, 1–7. https://doi.org/10.1109/ICCCNT51525.2021.9579611

[2] Bushra Aloraini, Meiyappan Nagappan, Daniel M. German, Shinpei Hayashi, and Yoshiki Higo. 2019. An empirical study of security warnings from static application security testing tools. *Journal of Systems and Software* 158 (12 2019), 110427. https://doi.org/10.1016/j.jss.2019.110427

[3] Nathaniel Ayewah and William Pugh. 2010. The Google FindBugs fixit. *Proceedings of the 19th international symposium on Software testing and analysis*, 241–252. https://doi.org/10.1145/1831708.1831738

[4] Daniel Dalalana Bertoglio and Avelino Francisco Zorzo. 2017. Overview and open issues on penetration test. *Journal of the Brazilian Computer Society* 23 (12 2017), 2. Issue 1. https://doi.org/10.1186/s13173-017-0051-1

[5] Wachiraphan Charoenwet, Patanamon Thongtanunam, Van-Thuan Pham, and Christoph Treude. 2024. An Empirical Study of Static Analysis Tools for Secure Code Review. arXiv:2407.12241 [cs.SE] https://arxiv.org/abs/2407.12241

[6] CrowdStrike. 2024. Technical details: Falcon update for windows hosts: CrowdStrike. https://www.crowdstrike.com/blog/falcon-update-for-windows-hosts-technical-details/

[7] Daniele Granata, Massimiliano Rak, and Giovanni Salzillo. 2022. MetaSEnD: A Security Enabled Development Life Cycle Meta-Model. *Proceedings of the 17th International Conference on Availability, Reliability and Security*, 1–10. https://doi.org/10.1145/3538969.3544463

[8] Daniel Graziotin, Fabian Fagerholm, Xiaofeng Wang, and Pekka Abrahamsson. 2017. Consequences of Unhappiness while Developing Software. *2017 IEEE/ACM 2nd International Workshop on Emotion Awareness in Software Engineering (SEmotion)*, 42–47. https://doi.org/10.1109/SEmotion.2017.5

[9] Zhaoqiang Guo, Tingting Tan, Shiran Liu, Xutong Liu, Wei Lai, Yibiao Yang, Yanhui Li, Lin Chen, Wei Dong, and Yuming Zhou. 2023. Mitigating False Positive Static Analysis Warnings: Progress, Challenges, and Opportunities. *IEEE Transactions on Software Engineering* 49 (12 2023), 5154–5188. Issue 12. https://doi.org/10.1109/TSE.2023.3329667

[10] Sarra Habchi, Xavier Blanc, and Romain Rouvoy. 2018. On adopting linters to deal with performance concerns in Android apps. *Proceedings of the 33rd ACM/IEEE International Conference on Automated Software Engineering*, 6–16. https://doi.org/10.1145/3238147.3238197

[11] Brittany Johnson, Yoonki Song, Emerson Murphy-Hill, and Robert Bowdidge. 2013. Why don't software developers use static analysis tools to find bugs? *2013 35th International Conference on Software Engineering (ICSE)*, 672–681. https://doi.org/10.1109/ICSE.2013.6606613

[12] Nenad Jovanovic, Christopher Kruegel, and Engin Kirda. 2006. Pixy: a static analysis tool for detecting Web application vulnerabilities. *2006 IEEE Symposium on Security and Privacy (SP'06)*, 6 pp.–263. https://doi.org/10.1109/SP.2006.29

[13] Barbara Kitchenham, Stuart Charters, et al. 2007. Guidelines for performing systematic literature reviews in software engineering version 2.3. *Engineering* 45, 4ve (2007), 1051.

[14] Kathleen Lindlan, Janice Cuny, Allen D. Malony, Sameer Shende, Bernd Mohr, Reid Rivenburgh, and Craig Rasmussen. 2000. A Tool Framework for Static and Dynamic Analysis of Object-Oriented Software with Templates. *ACM/IEEE SC 2000 Conference (SC'00)*, 49–49. https://doi.org/10.1109/SC.2000.10052

[15] V Benjamin Livshits and Monica S Lam. 2005. Finding security vulnerabilities in java applications with static analysis. *Proceedings of the 14th Conference on USENIX Security Symposium - Volume 14*, 18.

[16] Tamara Lopez, Helen Sharp, Arosha Bandara, Thein Tun, Mark Levine, and Bashar Nuseibeh. 2023. Security Responses in Software Development. *ACM Transactions on Software Engineering and Methodology* 32 (7 2023), 1–29. Issue 3. https://doi.org/10.1145/3563211

[17] Linghui Luo, Martin Schäf, Daniel Sanchez, and Eric Bodden. 2021. IDE support for cloud-based static analyses. *Proceedings of the 29th ACM Joint Meeting on European Software Engineering Conference and Symposium on the Foundations of Software Engineering*, 1178–1189. https://doi.org/10.1145/3468264.3468535

[18] Diego Marcilio, Carlo A. Furia, Rodrigo Bonifacio, and Gustavo Pinto. 2019. Automatically Generating Fix Suggestions in Response to Static Code Analysis Warnings. *2019 19th International Working Conference on Source Code Analysis and Manipulation (SCAM)*, 34–44. https://doi.org/10.1109/SCAM.2019.00013

[19] Jose Andre Morales, Thomas P. Scanlon, Aaron Volkmann, Joseph Yankel, and Hasan Yasar. 2020. Security impacts of sub-optimal DevSecOps implementations in a highly regulated environment. *Proceedings of the 15th International Conference on Availability, Reliability and Security*, 1–8. https://doi.org/10.1145/3407023.3409186

[20] Marcus Nachtigall, Michael Schlichtig, and Eric Bodden. 2022. A large-scale study of usability criteria addressed by static analysis tools. *Proceedings of the 31st ACM SIGSOFT International Symposium on Software Testing and Analysis*, 532–543. https://doi.org/10.1145/3533767.3534374

[21] Martin Odermatt, Diego Marcilio, and Carlo A. Furia. 2022. Static Analysis Warnings and Automatic Fixing: A Replication for C Projects. *2022 IEEE International Conference on Software Analysis, Evolution and Reengineering (SANER)*, 805–816. https://doi.org/10.1109/SANER53432.2022.00098

[22] Eric O'Donoghue, Ann Marie Reinhold, and Clemente Izurieta. 2024. Assessing Security Risks of Software Supply Chains Using Software Bill of Materials. *2nd International Workshop on Mining Software Repositories for Privacy and Security, (SANER 2024).*

[23] Yuanyuan Pan. 2019. Interactive Application Security Testing. *2019 International Conference on Smart Grid and Electrical Automation (ICSGEA)*, 558–561. https://doi.org/10.1109/ICSGEA.2019.00131

[24] Roshan Namal Rajapakse, Mansooreh Zahedi, and Muhammad Ali Babar. 2021. An Empirical Analysis of Practitioners' Perspectives on Security Tool Integration into DevOps. *Proceedings of the 15th ACM / IEEE International Symposium on Empirical Software Engineering and Measurement (ESEM)*, 1–12. https://doi.org/10.1145/3475716.3475776

[25] Ann Marie Reinhold, Travis Weber, Colleen Lemak, Derek Reimanis, and Clemente Izurieta. 2023. New Version, New Answer: Investigating Cybersecurity Static-Analysis Tool Findings. *2023 IEEE International Conference on Cyber Security and Resilience (CSR)*, 28–35. https://doi.org/10.1109/CSR57506.2023.10224930

[26] Markus Schnappinger, Mohd Hafeez Osman, Alexander Pretschner, and Arnaud Fietzke. 2019. Learning a Classifier for Prediction of Maintainability Based on Static Analysis Tools. *2019 IEEE/ACM 27th International Conference on Program Comprehension (ICPC)*, 243–248. https://doi.org/10.1109/ICPC.2019.00043

[27] Tyler W. Thomas, Madiha Tabassum, Bill Chu, and Heather Lipford. 2018. Security During Application Development. *Proceedings of the 2018 CHI Conference on Human Factors in Computing Systems*, 1–12. https://doi.org/10.1145/3173574.3173836

[28] Zachary Wadhams, Ann Marie Reinhold, and Clemente Izurieta. 2024. Automating Static Analysis Through CI/CD Pipeline Integration. *2nd International Workshop on Mining Software Repositories for Privacy and Security, (SANER 2024).*
[29] Marvin Wyrich, Andreas Preikschat, Daniel Graziotin, and Stefan Wagner. 2021. The Mind Is a Powerful Place: How Showing Code Comprehensibility Metrics Influences Code Understanding. *2021 IEEE/ACM 43rd International Conference on Software Engineering (ICSE)*, 512–523. https://doi.org/10.1109/ICSE43902.2021.00055
[30] Jing Xie, Heather R. Lipford, and Bill Chu. 2011. Why do programmers make security errors? *2011 IEEE Symposium on Visual Languages and Human-Centric Computing (VL/HCC)*, 161–164. https://doi.org/10.1109/VLHCC.2011.6070393
[31] Jinqiu Yang, Lin Tan, John Peyton, and Kristofer A Duer. 2019. Towards Better Utilizing Static Application Security Testing. *2019 IEEE/ACM 41st International Conference on Software Engineering: Software Engineering in Practice (ICSE-SEIP)*, 51–60. https://doi.org/10.1109/ICSE-SEIP.2019.00014